\documentclass[prd,superscriptaddress,onecolumn,showkeys]{revtex4}
\usepackage{eurosym}
\usepackage{amsfonts}
\usepackage{array}
\usepackage{amsthm}
\usepackage{bm}
\usepackage{palatino}
\usepackage{mathpazo}
\usepackage{amssymb}
\usepackage{eurosym}
\usepackage{amsmath}
\usepackage{epsfig}
\usepackage{graphics}
\usepackage{changes}
\usepackage{color}
\usepackage{graphicx}
\usepackage[colorlinks=true,
            linkcolor=blue,
           urlcolor=black,
           citecolor=black]{hyperref}

\def\be{\begin{equation}}
\def\ee{\end{equation}}
\def\beq{\begin{eqnarray}}
\def\eeq{\end{eqnarray}}

\def\bes{\begin{eqnarray}}
\def\ees{\end{eqnarray}}

\begin{document}

\title{Tunneling and thermodynamics evolution of the magnetized Ernst-like black hole}

\author{Riasat Ali}
\email{riasatyasin@gmail.com}
\affiliation{Department of Mathematics, GC
University Faisalabad Layyah Campus, Layyah-31200, Pakistan}

\author{Zunaira Akhtar}
\email{zunaira.math.pu@gmail.com}
\affiliation{Department of Mathematics, University of the Punjab, Quaid-e-Azam Campus, Lahore 54590, Pakistan}

\author{Kazuharu Bamba}
\email{bamba@sss.fukushima-u.ac.jp}
\affiliation{Faculty of Symbiotic Systems Science
Fukushima University, Fukushima 960-1296, Japan}

\author{M. Umar Khan}
\email{umar\_khan@comsats.edu.pk}
\affiliation{Center for Advanced Studies in Telecommunications, COMSATS University Islamabad, Pakistan.}

\begin{abstract} 
We investigate the tunneling phenomenon of particles through the horizon of a magnetized Ernst-like black hole. We employ the modified Lagrangian equation with the extended uncertainty principle for this black hole. We determine a tunneling rate and the related Hawking temperature for this black hole by using the WKB approach in the field equation. In addition, we examine the graph behavior of the Hawking temperature in relation to the black hole event horizon. We explore the stability analysis of this black hole by taking into account the impact of quantum gravity on Hawking temperatures. The temperature for a magnetized Ernst-like black hole rises as the correction parameter is decreased. Moreover, we analyze the thermodynamics quantities such as Hawking temperature, heat capacity and Bekenstein entropy by using the different approach. We obtain the corrected entropy to study the impact of logarithmic corrections on the different thermodynamic quantities. It is shown that these correction terms makes the system stable under thermal fluctuations.
\end{abstract}

\date{\today}

\maketitle

\section{Introduction}

The black holes (BHs) is a real physical object, due to its powerful gravitational pull
that has the capacity to absorb all forms of energy from its surroundings. The general
theory of relativity states that a BH attracts all kinds of particles that interact
with the event horizon. By taking quantum effects into account against the background
of curled spacetime, Hawking explained in 1974 that a BH acts as a black body and radiates
particles in the form of radiation through its horizon. These particles are referred \cite{1} to
as "\textit{Hawking radiation}" and have a specific temperature known as "\textit{Hawking temperature}".
One of the efficient approaches is \textit{quantum tunneling} for exploring the Hawking radiation phenomenon \cite{2,3,4}.
This process based on the formation of electron-positron pairs, which involves an electric field.
A pair generated just outside or inside the horizon may exist with zero total energy, and one pair may have the potential to tunnel to the inside of the horizon because it may be a
well-known fact that once particles cross the outside horizon \cite{5}.

The tunneling phenomenon states that by beginning from outer horizon and extending to infinity,
particles are considered to follow the usually classical forbidden trajectories \cite{6,7}.
The emission of particles in the form of radiation, which leads lose of BH mass,
can be used to study the evaporation of BH. A BH evaporates, shrinks, and finally disappears
when it loses more radiations than it obtains through many other techniques.
This phenomenon influences a BH's charge, mass, and angular momentum, which are all thermodynamic properties.
These particles move in accordance with the radial null geodesics frame.
While the particles are assumed to be real for ingoing geodesics, they must be imaginary for outgoing geodesics.
Its only actual particle that can exist inside the horizon has a speed that is
not exactly or equal to the light speed.
There are two basic methods for determining the imaginary portion of the classical action: the
\textit{null geodesic approach} and the \textit{Hamilton-Jacobi approach}.
The first one was implemented by Parikh and his collaborators \cite{8}, and
the second was presented by Srinivasan and Padmanabhan \cite{9}.
Wentzel, Kramers, and Brillouin (WKB) \cite{10} used an approximation known as the
WKB approximation to develop a connection between classical and quantum theory.
In order to determine the Hawking temperature for various BHs and wormholes \cite{11,12,15,21,22,23,24,27,28,29,30,32,33,35,37,39,42,43,44,45,46,47,48,49,50,51,53,54,56,57,58,59,60,61,63},
many researchers examined the bosonic tunnelng  of particles, spin-3/2, spin-2, and fermion tunneling particles.
Jian and Bing-Bing \cite{64} investigated the fermion tunneling of particles from charged and uncharged BHs.
Yale examined \cite{65} the precise Hawking temperatures for the fermion, scalar, and boson particles
that emerge from BHs without influences of back-reaction. The quantum tunneling mechanism was used by
Sharif and Javed \cite{66} to study the Hawking radiation for various types of BHs and to derive the
tunneling probability and corresponding temperatures. The fermion tunneling behaviour of particles
from wormholes was estimated by the same authors in \cite{67}. They also discussed how
fermion particles behaved when charged, rotating, accelerating BHs with the $NUT$ parameter \cite{68} were present.

In recent years, the universe has provided some logical evidence in favour of the
Einstein gravity theory. The conservation energy and momentum law is explained
by this theory of gravity. The Einstein gravity theory, however, can also be seen
as a specific case of several modified theories of gravity.
The majority of scientists have been investigating several
specific theories and developing some modified gravity theories.
It is possible to discuss about the corrected quantum thermodynamical features of BH \cite{68a, 69}
by considering the effects of the generalized uncertainty principle (GUP).
Chen \cite{70} analyzed into consideration the correction values in accordance with this condition,
which also fulfills the GUP relation necessity and also taken into consideration first order terms
of correction for Hawking temperature corrections. The GUP concept has been applied to numerous BHs.
For Kerr, Kerr-Newman and Reissner-Nordstrom BHs \cite{71}, the tunneling process makes a significant
contribution to BH physics. Jiang \cite{72} computed Dirac particle tunneling and examined the Hawking
radiation for black ring. In order to examine the Hawking temperature of particles, Ali et al.
\cite{78} used the $WKB$ approximation and the Lagrangian field~equation
with/without GUP to study the vector particles tunneling from various BHs or black
ring with a different~parameters.

Ghaderi and Malakolkalami \cite{82} studied the thermodynamics of the Schwarzschild BH and derived the temperature, heat capacity and mass associated with entropy. In order to increase the accuracy of their findings, many authors have examined the thermodynamics of different kinds of BHs under the effects of quantum gravity and converted their results into Schwarzschild Hawking temperatures \cite{T4,T5}. Within the framework of Weyl corrections, Sharif and Zunaira \cite{T7,T8} have studied the quasi-normal mode, thermodynamics, and thermal fluctuations of charged BHs. It should be noted that the system seems to be more unstable for small BH radii when first-order corrections are present. Through the use of the Hessian matrix and heat capacity, they have also examined the stability of the system. Additionally, these authors have looked at the local and global stability of the Newman-Unti-Tamburino BH's thermodynamics and phase transition in the presence of charged, accelerating, and rotating pairs \cite{T9,T10}. Our paper's objectives are to analyze the magnetised Ernst-like BH $T_H$ under the influence of the GUP parameter in the background
of the Hamilton-Jacobi method, and describe a comparison of our new findings
with previous research. To calculate the quantum $T'_{H}$
for magnetised Ernst-like BH with GUP effects as well as to examine magnetised
Ernst-like BH stability in the presence of quantum gravity effects.

This paper is arranged in the following way: Section \textbf{II}, comprises a brief presentation about the metric of
magnetised Ernst-like BH and also investigate the $T'_{H}$ of BH under the influence
of GUP parameter. In the Sec. \textbf{III} presents the graph analysis of $T'_{H}$
w.r.t outer horizon. Sec. \textbf{IV} investigates the corrected entropy to see the influences of thermal fluctuations on the considered geometry. Finally, section \textbf{V}, analyzed the result and discussion.

\section{Hawking Radiation from Magnetised
Ernst-Like Black Hole}
We examine the motion of boson particles in the horizon of a magnetised
Ernst like BH while taking into account its surrounding magnetic field.
We demonstrate how its magnetic field can affect the Hawking radiation.
The four-dimensional Ernst metric \cite{c1, c2} has the following form

\begin{equation}
ds^{2}=-A^{2}(r, \theta)B(r)dt^2+A^{2}(r, \theta)\frac{1}{B(r)}dr^2+
A^{2}(r, \theta)r^2 d\theta^2+\frac{1}{A^{2}(r, \theta)}r^2sin^2\theta d\phi^2,\label{st1}
\end{equation}
with
\begin{eqnarray}
A(r, \theta)&=&1+r^{2}B^{2}\sin^{2}\theta,\nonumber\\
B(r)&=&1-2Mr^{-1},\nonumber
\end{eqnarray}
where the $B$ parameter is a magnetic field. The metric is not asymptotically flat
and is not spherically symmetric since there is a strong magnetic field present.
The spacetime in equation (\ref{st1}) can be rewritten as
\begin{equation}
ds^{2}=-Fdt^{2}+Gdr^{2}+Hd\theta^{2}
+Id\phi^{2},\label{st2}
\end{equation}
where
\begin{eqnarray}
F&=&A^{2}(r, \theta)B(r),\nonumber\\
G&=&\frac{A^{2}(r, \theta)}{B(r)},\nonumber\\
H&=& A^{2}(r, \theta)r^2,\nonumber\\
I&=& \frac{r^2sin^2\theta}{A^{2}(r, \theta)}.\nonumber
\end{eqnarray}
We examine the boson tunnel that forms around a magnetized Ernst-like BH while
taking into account the magnetic field its surroundings.~We demonstrate how the
BH magnetic field can affect particle tunneling.
The generally Lagrangian field equation with GUP parameter can be written \cite{33, T5} as
\begin{eqnarray}
&&\partial_{\mu}(\sqrt{-\textbf{g}}\Phi^{\nu\mu})+\sqrt{-\textbf{g}}\frac{m^2}{\hslash^2}\Phi^{\nu}+
\sqrt{-\textbf{g}}\frac{i}{\hslash}eA_{\mu}\Phi^{\nu\mu}
+\sqrt{-\textbf{g}}\frac{i}{\hslash}\Phi_{\mu}eF^{\nu\mu}
+\hslash^{2}\partial_{0}\partial_{0}\partial_{0}(\sqrt{-\textbf{g}}\textbf{g}^{00}\Phi^{0\nu})\nonumber\\
&&-\alpha \hslash^{2}\partial_{i}\partial_{i}\partial_{i}(\sqrt{-\textbf{g}}\textbf{g}^{ii}\Phi^{i\nu})=0,\label{L}
\end{eqnarray}
here $\Phi^{\nu\mu}$, $\textbf{g}$ and  $m$ show the anti-symmetric tensor, determinant of coefficient matrix
and particle mass. The non-zero elements of anti-symmetric tensor $\Phi_{\nu\mu}$ can be computed as
\begin{equation}
\Phi_{\nu\mu}=(1-\hslash^2\alpha{\partial_\nu}^2)\partial_{\nu}\Phi_{\mu}-
(1-\hslash^2\alpha{\partial_\mu}^2)\partial_{\mu}\Phi_{\nu}+(1-\hslash^2\alpha{\partial_\nu}^2)
\frac{i}{\hslash}eA_{\nu}\Phi_{\mu}-(1-\hslash^2\alpha{\partial_\mu}^2)\frac{i}{\hslash}eA_{\mu}\Phi_{\nu},\label{q}\nonumber
\end{equation}
and
\begin{equation}
F_{\nu\mu}=D_{\nu} A_{\mu}-D_{\mu} A_{\nu},~~~D_{o}=\Big(1
+\hslash^2\alpha \textbf{g}^{00}D^2_{o}\Big)D_{o},~~~D_{i}=\Big(1-\hslash^2\alpha \textbf{g}^{ii}D^2_{i}\Big)D_{i},\nonumber
\end{equation}
here $\alpha$, $A_{\mu},~e~$ and $D_{\mu}$ are GUP parameter (quantum gravity), BH potential,
charge of particle and covariant derivative, respectively.
The elements of $\Phi^{\mu}$ and $\Phi^{\mu\nu}$ is evaluated as
\begin{eqnarray*}
\Phi^{0}&=&\frac{\Phi_{0}}{F},~~~\Phi^{1}=
\frac{\Phi_{1}}{G},~~~\Phi^{2}=\frac{\Phi_{2}}{H},~~~
\Phi^{3}=\frac{\Phi_{3}}{I},\\
\Phi^{01}&=&\frac{\Phi_{01}}{FG},~~~\Phi^{02}=
\frac{\Phi_{02}}{FH},~~~
\Phi^{03}=\frac{\Phi_{03}}{FI},\\
\Phi^{12}&=&\frac{\Phi_{12}}{GH},~~
\Phi^{13}=\frac{\Phi_{13}}{GI},~~
\Phi^{23}=\frac{\Phi_{23}}{HI}.
\end{eqnarray*}
The WKB approximation is expressed in \cite{78}, i.e.,
\begin{equation}
\Phi_{\nu}=c_{\nu}\exp[\frac{i}{\hslash}K_{0}(t,r,\theta,\phi)+
\sum_{i=1}^{i=n} \hslash^{i}K_{i}(t,r,\theta,\phi)].\label{wkb1}
\end{equation}
Substituting the Eq. (\ref{wkb1}) into the Eq. (\ref{L}),with $i=1,2,3,...$ neglecting the
higher terms and get the set of Eqs. below in appendix A.
We can take the variables separation,
\begin{equation}
K_{0}=-\xi t
+W(r)+j\phi+v(\theta),\label{RRA}
\end{equation}
with $\xi=(E-j\Omega)$ and $\Omega$, $j$ and $E$ represent angular particle velocity, angular momentum and energy.
respectively. Here, $W(r)$ and $v(\theta)$ are two arbitrary functions.~The matrix equation can be get from the Eqs.~(\ref{31})--(\ref{61}),
\begin{equation*}
S(c_{0},c_{1},c_{2},c_{3})^{T}=0,
\end{equation*}
which provides "$S$" is a $'4 \times 4'$ order matrix, and its elements are as follows:
\begin{eqnarray}
S_{00}&=&\frac{\dot{W}^{2}+\alpha\dot{W}^{4}}{G}-\frac{j^{2}+\alpha j^{4}}{H}
+\frac{\dot{v}^{3}+\alpha\dot{v}^{4}}{I}-m^{2},~~~
S_{01}=-\frac{\dot{W}\xi+\alpha\dot{W}\xi^{3}}{G}
+\frac{\dot{W}eA_{0}+\alpha\dot{W}eA_{0}\xi^{2}}{G},\nonumber\\
S_{02}&=&-\frac{\xi j+\alpha\xi j}{H}+
\frac{eA_{0}j+\alpha\xi^{2}eA_{0}j}{H},~~~
S_{03}=-\frac{\dot{v}\xi+\alpha\dot{v}\xi^{3}}
{I}+\frac{eA_{0}\dot{v}+\alpha eA_{0}\dot{v}\xi^{2}}{I},\nonumber\\
S_{10}&=&\frac{\xi\dot{W}+\alpha\xi\dot{W}^{3}}
{F}-\frac{eA_{0}\dot{W}+\alpha eA_{0}\dot{W}^{3}}{F},\nonumber\\
S_{11}&=& \frac{\xi^{2}+\alpha\xi^{4}}{F}+
\frac{\xi eA_{0}-\alpha\dot{W}
\xi eA_{0}}{F}
-\frac{{j}^{2}-\alpha {j}^{4}}{H}-\frac{\dot{v}^{2}-\alpha\dot{v}^{4}}{I}
-m^{2}-\frac{1}{F(r)}eA_{0}[\xi+\alpha\xi^{3}
-eA_{0}-\alpha eA_{0}\dot{W}^{2}],\nonumber\\
S_{12}&=&\frac{\dot{W}j+\alpha\dot{W}^{3}j}{H},~~~~
S_{13}=\frac{\dot{v}\dot{W}+\dot{v}\alpha\dot{W}^{3}}{I},~~~
S_{20}=-\frac{j\xi+\alpha j^{3}\xi}{F}
-eA_{0}\frac{j+\alpha j^{3}}{F},~~~
S_{21}=\frac{\dot{W}j+\alpha\dot{W}j^{3}}{G},\nonumber\\
S_{22}&=&-\frac{1}{F}[-\xi^{2}-\alpha \xi^{4}+eA_{0}
\xi+eA_{0}\alpha\xi^{3}]
-\frac{\dot{W}^{2}+\alpha\dot{W}^{4}}{G}
-\frac{\dot{v}^{2}+\alpha\dot{v}^{4}}{I}-m^{2},~~
S_{23}=\frac{j\dot{v}+\alpha j^{3}\dot{v}}{I},\nonumber\\
S_{30}&=&\frac{-1}{F}[\xi\dot{v}+\alpha\xi\dot{v}^{3}]
+\frac{eA_{0}\dot{v}+eA_{0}\alpha\dot{v}^{3}}{F},~~
S_{31}=\frac{-\dot{W}\dot{v}-\alpha\dot{W}\dot{v}^{3}}{G},~~
S_{32}=\frac{-j\dot{v}-\alpha j\dot{v}^{3}}{H},\nonumber\\
S_{33}&=&-\frac{1}{F}[\xi^{2}+\alpha\xi^{4}-
\xi eA_{0}-\alpha\xi eA_{0}\dot{v}^{3}]+
\frac{\dot{W}^{2}+\alpha\dot{W}^{4}}{G}-\frac{j^{2}+\alpha j^{4}}{H}
+\frac{eA_{0}}{F}[\xi+\alpha\xi^{3}
-eA_{0}-\alpha eA_{0}\dot{v}^{3}]-m^{2},\nonumber
\end{eqnarray}
where $\dot{W}=\partial_{r}K_{0}$,
$\dot{v}=\partial_{\theta}K_{0}$ and
 $j=\partial_{\phi}K_{0}$.
Solving these equations for trivial solution $\mid\textbf{S}\mid=0$ in the following result:
\begin{equation}\label{w1}
ImW^{\pm}=\pm \int\sqrt{\frac{(\xi-eA_{0})^{2}+X_{1}(1+\frac{X_{2}}{X_{1}}\alpha)}{G^{-1}}}dr,
\end{equation}
where the outgoing and ingoing particles are indicated by $+$ and $-$, respectively.
The description of the function $'X_1'$ and $'X_2'$ are
\begin{eqnarray}
X_{1}&=&\frac{j^{2}}{H}\nonumber\\
X_{2}&=&\frac{\alpha\xi^{4}}{F}-\frac{\alpha\xi eA_{0}
\dot{v}^{3}}{F}-
\frac{\alpha\dot{W}^{4}}{G}+\alpha\frac{j^{4}}{H}
-\frac{eA_{0}}{F}[\alpha\xi^{3}-\alpha
eA_{0}\dot{v}^{3}] +m^{2}\nonumber,
\end{eqnarray}
indicate the angular velocity at
the BH horizon. Equation (\ref{w1}) is integrated around the pole to yield
\begin{equation}
ImW^{\pm}
=\pm i\pi\frac{(\xi-eA_{0})}{2k(r_{+})}[1+Y\alpha],
\end{equation}
with $k(r_{+})$ and $Y$ are the surface gravity and arbitrary parameter, respectively and the
4-dimensional magnetised Ernst-like BH surface gravity is defined by
\begin{equation}
k(r_{+})=\frac{8M}{r_{+}}B^{2}\sin^{2}\theta\Big(r_{+}^{2}B^{2}\sin^{2}\theta+1\Big).
\end{equation}
For boson vector particles, the tunneling probability $\Gamma(ImW^{+})$ is given by
\begin{equation}
\Gamma(ImW^{+})=\frac{\Gamma_{[emission]}}{\Gamma_{[absorption]}}
=\exp\Big[-\pi\frac{(\xi-eA_{0})r_{+}}{4MB^{2}\sin^{2}\theta\Big(r_{+}^{2}B^{2}\sin^{2}\theta+1\Big)}\Big]
[1+Y\alpha].\label{TUN}
\end{equation}

By comparing the $\Gamma(imW^{+})$ with the Boltzmann formula, we can now compute the Hawking
temperature $\Gamma(ImW^{+})\approx e^{-(\xi-eA_{0})/T'_{H}}$, we get
\begin{equation}\label{TH}
T'_{H}=\frac{4MB^{2}\sin^{2}\theta\Big(r_{+}^{2}B^{2}\sin^{2}\theta+1\Big)}{\pi r_{+}}
[1-Y\alpha].
\end{equation}
The radial coordinate at the outside horizon, the BH mass, the GUP parameter and the magnetic field affect the Hawking temperature.

\section{Analysis of Hawking temperature}

We examine the evolution of
temperature for different correction parameter  values in left hand side of figure 1 while~considering
constant values for other parameters. After reaching a maximum height,
the $T'_{H}$ in the domain of $0\leq r \leq 1$ decreases. The $T'_{H}$ is
reflected by the maximum temperature with a non-zero horizon.
This $T'_{H}$ physical behavior suggests that the positive range of BH stability is present. It is also worth noting that the
$T'_H$ increases as the correction value $\alpha$ decreases.
On the right hand side of \textit{Fig.} 1, demonstrates how, depending
on the magnetic field levels, the $T'_{H}$~eventually~increases
from a height and then gradually decreases. The decrease in $T'_H$
with raising horizon demonstrates BH stable condition in the decrease domain
of $0\leq r \leq 1$. It is obvious that as the BH magnetic field increases,
the $T'_H$ value increases.

\begin{center}
\includegraphics[width=7cm]{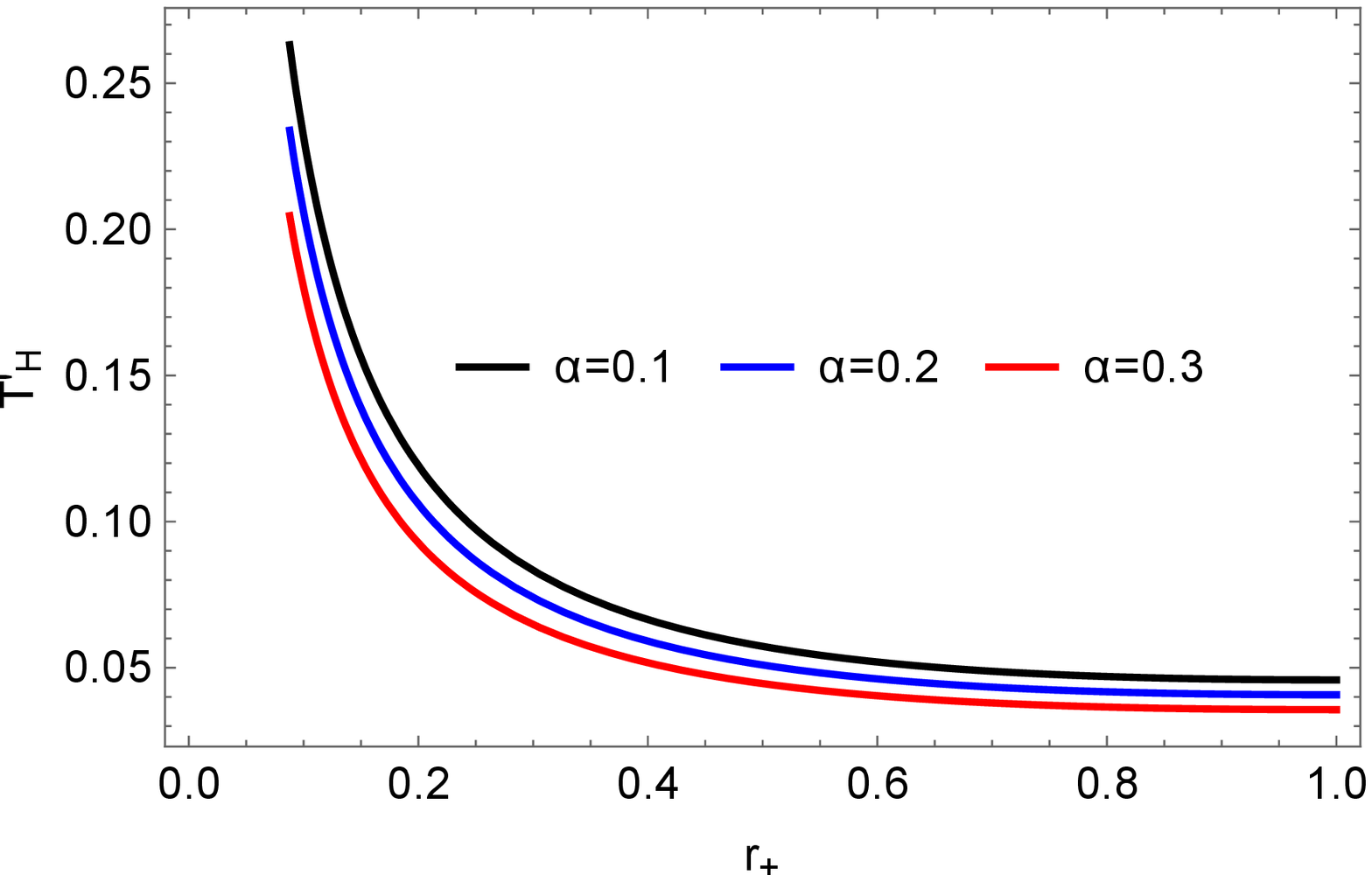}\includegraphics[width=7cm]{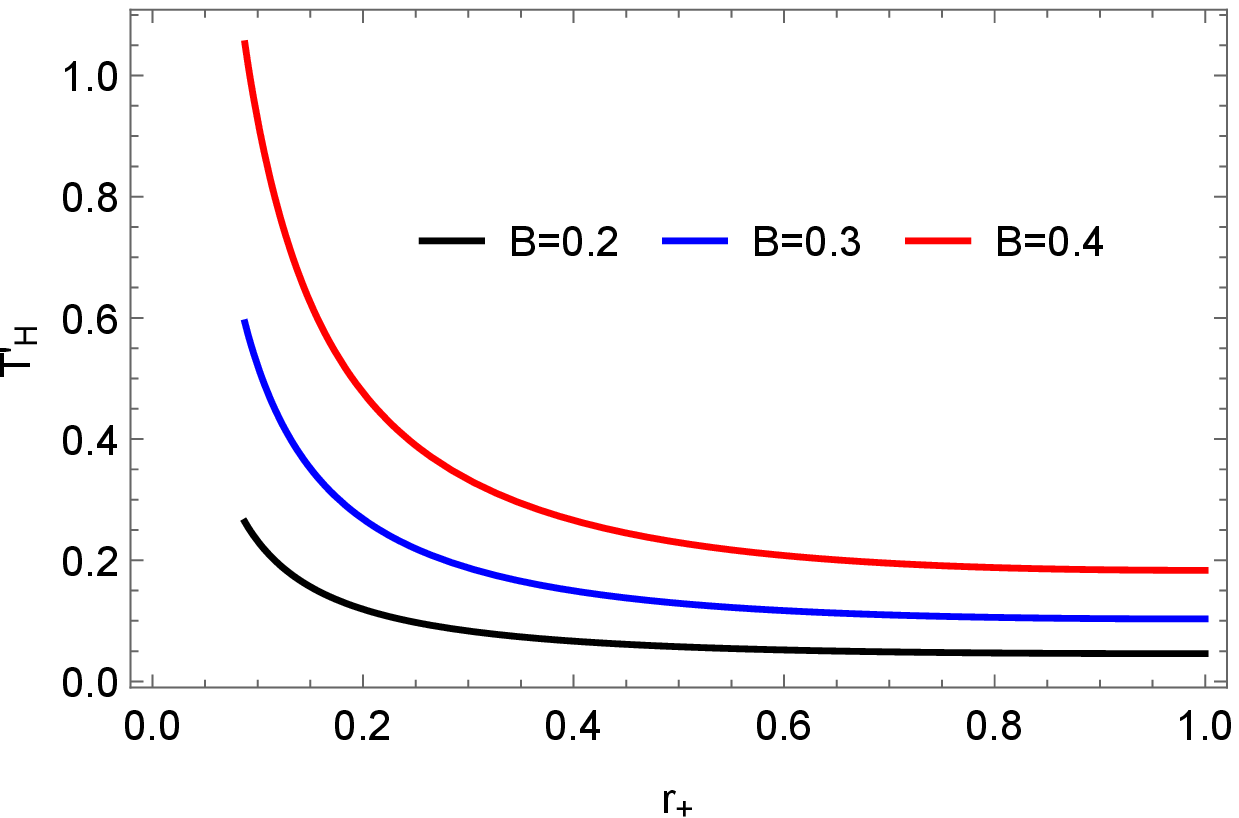}\\
{Figure 1: $T'_{H}$ via horizon $r_{+}$ for $M=0.5$, $Y=1$ and $B=0.2$. Left $\alpha=0.1$ (black), $\alpha=0.2$ (blue), $\alpha=0.3$ (red).
Right $M=0.5$, $Y=1$, $\alpha=0.1$, $B=0.2$ (black), $B=0.3$ (blue), $B=0.4$ (red).}
\end{center}

\section{Thermal Fluctuations}
Thermal fluctuations plays an important role on the study of BH thermodynamics. With the concept of Euclidean quantum gravity, the temporal coordinates shifts towards complex plan. To check, the effects of these correction in
entropy, we have need to find usual entropy of the given system with the help of first law of thermodynamics by the
formula
\begin{eqnarray}
S=\int T^{-1}\frac{\partial m}{\partial r_{+}}dr_{+},
\end{eqnarray}
by inserting of value Hawking temperature (\ref{TH}), the entropy expression takes the form
\begin{eqnarray}\label{s}
S=\Big(B^2 r_{+}^3  \sin^{2}{\theta} \Big(B^2 r_{+} (2 r_{+}-3 M)\sin^{2}{\theta}+2\Big)+M\Big)(2 \pi  r_{+}^2)^{-1}
\end{eqnarray}
To check the corrected entropy along these thermal fluctuations, the partition function is $Z(\mu)$ in terms of  density of states $\eta(E)$ is given as 
\begin{equation}
Z(\mu)=\int_{0}^{\infty} \exp(-\mu E)\eta(E)dE,
\end{equation}
where $T_{+}=\frac{1}{\mu}$ and E is the average energy of thermal radiations. With the help of Laplace inverse transform, the expression of density takes the form
\begin{equation}
\rho(E)=\frac{1}{2\pi
i}\int_{\mu_{0}-i\infty}^{\mu_{0}+i\infty} Z(\mu)
\exp(\mu E) d\mu=\frac{1}{2\pi
i}\int_{\mu_{0}-i\infty}^{\mu_{0}+i\infty}
\exp(\tilde{S}(\mu))d\mu,
\end{equation}
where $\tilde{S}(\mu)=\mu E+ \ln Z(\mu)$ represents the corrected entropy of the considered system which is dependent on Hawking temperature. Moreover, the expression of corrected entropy gets modified with the help of steepest decent method,
\begin{equation}
\tilde{S}(\mu)=S+\frac{1}{2}(\mu-\mu_{0})^{2}
\frac{\partial^{2}\tilde{S}(\mu)}{\partial
\mu^{2}}\Big|_{\mu=\mu_{0}}+\text{higher-order terms}.
\end{equation}
Using the conditions $\frac{\partial
\tilde{S}}{\partial\mu}=0$ and $\frac{\partial^{2}
\tilde{S}}{\partial\mu^{2}}>0$, the corrected entropy relation under the first-order corrections modified. By neglecting higher order terms, the exact expression of entropy is expressed as 
\begin{equation}\label{4}
\tilde{S}=S-\mho \ln(ST^{2}),
\end{equation}
where $\mho$ is called correction parameter, the usual entropy of considered system is attained by fixing $X=0$ that is without influence of these corrections. Furthermore, inserting the Eqs. (\ref{TH}) and (\ref{s}) into (\ref{4}), we have
\begin{eqnarray}
\tilde{S}&=&\Big(B^2r_{+}^3 \sin^{2}{\theta} \Big(B^2 r_{+} (2 r_{+}-3 M) \sin^{2}{\theta}+2\Big)+M)\Big(2 \pi r_{+}^2\Big)^{-1}-\mho  \log \Big(\Big(8 B^4 M^2 \sin^{2}{\theta}^2 (\alpha  Y-1)^2
\Big(B^2 r_{+}^2 \sin^{2}{\theta}+1\Big){}^2 \nonumber\\&\times&\Big(B^2 r_{+}^3 \sin^{2}{\theta} \Big(B^2 r_{+} (2 r_{+}-3 M) \sin^{2}{\theta}+2\Big)+M\Big)\Big)\Big(\pi ^3 r_{+}^4 \Big)^{-1}\Big).
\end{eqnarray}
\begin{center}
\includegraphics[width=8cm]{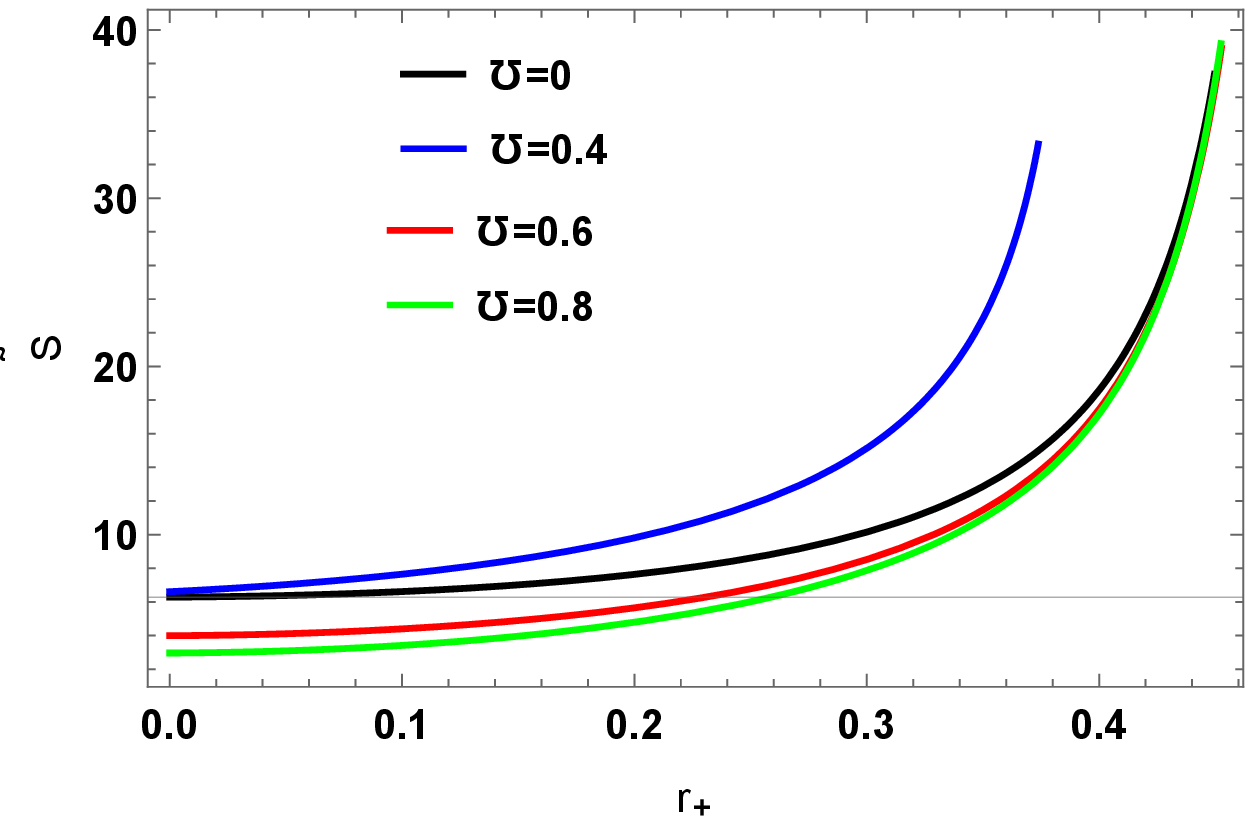}\\
{Figure 2: $\tilde{S}$ versus $r_{+}$ for $M=10,~B=0.5, \alpha=0.2$  and $Y=0.5$}.
\\
\end{center}
In the \textbf{Fig. 2}, the graph of corrected entropy is monotonically increasing throughout the considered domain. It is noted the graph (black) of usual entropy is increasing just for small value of horizon radius but corrected expression of energy is increasing smoothly. Thus, these corrections terms are more effective for small BHs. Now, using the expression of corrected entropy and check the other 
 other thermodynamic quantities under the effects of thermal fluctuations. In this way, the the Helmholtz energy ($F=-\int \tilde{S}dT$) leads to the form. We obtained expressions of Hawking temperature and corrected entropy are used to investigate various thermodynamical quantities under the impact of thermal fluctuations. The
Helmholtz free energy $(F=-\int \tilde{S} dT)$ of considered geometry in the thermal fluctuations presence can be computed as
\begin{eqnarray}
F&=&-\Big(2 B^2 M \sin^{2}{\theta} (\alpha  Y-1) \Big(B^2 r_+^2 \sin^{2}{\theta}-1\Big) \Big(r_+^2 \Big(B^2 r_+ \sin^{2}{\theta} \Big(B^2 r_+ \Big(2 r_+-3
   M\Big) \sin^{2}{\theta}+2\Big)-2 \pi  \mho  \log \nonumber\\&\times&\Big(\Big(8 B^4 M^2 \sin^{2}{\theta}^2 (\alpha  Y-1)^2 \Big(B^2 r_+^2 \sin^{2}{\theta}+1\Big){}^2
   \Big(B^2 r_+^3 \sin^{2}{\theta} \Big(B^2 r_+ \Big(2 r_+-3 M\Big) \sin^{2}{\theta}+2\Big)+M\Big)\Big)\Big(\pi ^3 r_+^4\Big)^{-1}\Big)\Big)\Big)\nonumber\\&\times&\Big(\pi ^2 r_+^4\Big)^{-1}.
\end{eqnarray}
 \begin{center}
\includegraphics[width=8cm]{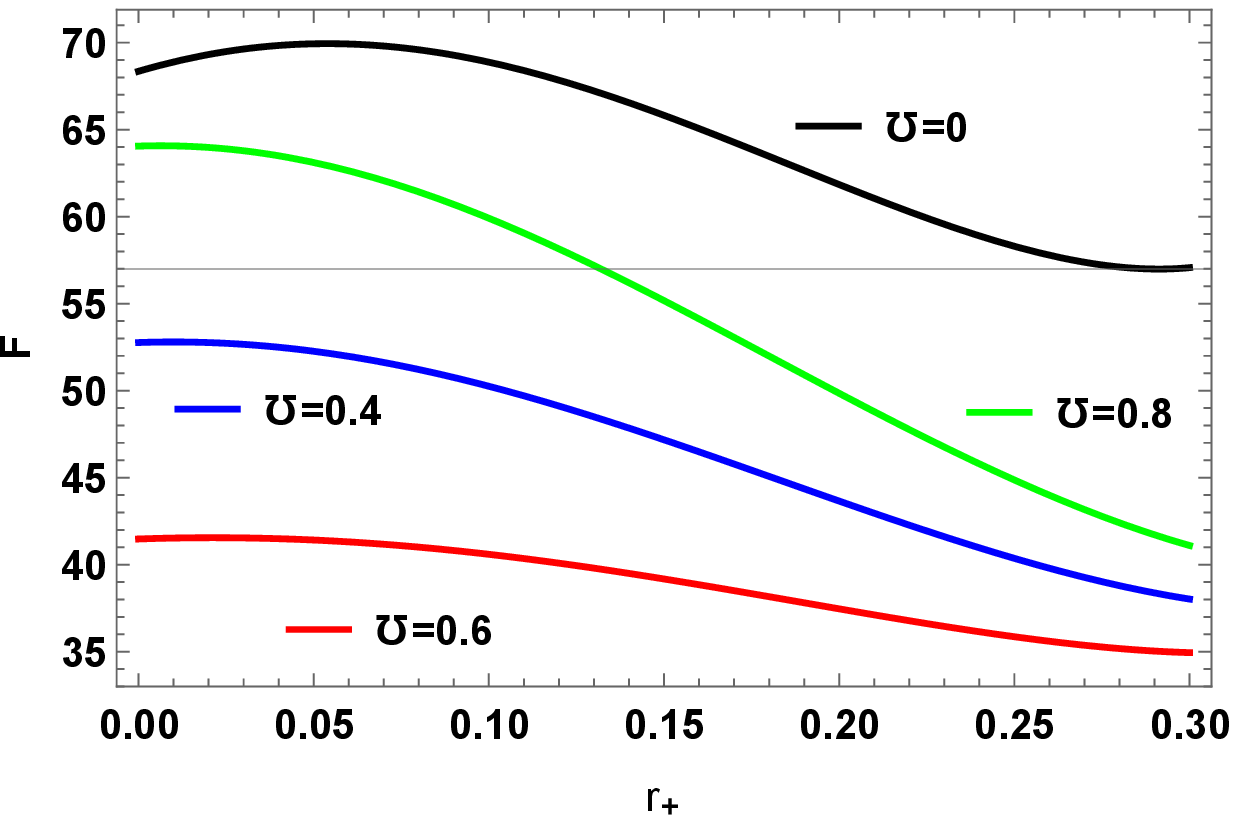}\\
{Figure 3: $F$ versus $r_{+}$ for  $M=10,~B=0.5, \alpha=0.2$  and $Y=0.5$}.
\\
\end{center}
The \textbf{Fig. 3} represents the graph of Helmholtz free energy with respect to horizon radius. It is observed that the behaviour of energy is gradually decreases for the different values of correction parameter $\mho$. While the graph of usual entropy shows opposite behaviour as the graph is increasing. This behaviour means that the considered system shifts its state towards equilibrium, thus, no more work can be extract from it. The expression of internal energy ($E=F+T\tilde{S}$) for the considered geometry is computed as 
\begin{eqnarray}
E&=&\Big(2 B^2 M \sin^{2}{\theta}  (\alpha  Y-1) \Big(r_+ \Big(B^2 r_+ \Big(r_++1\Big) \sin^{2}{\theta} +1\Big)-1\Big) \Big(r_+^2 \Big(B^2 r_+  \sin^{2}{\theta} \Big(B^2 r_+ \Big(2 r_+-3 M\Big) \sin^{2}{\theta} +2\Big)\nonumber\\&-&2 \pi  \mho  \log \Big(\Big(8 B^4 M^2 \sin^{2}{\theta} ^2 (\alpha  Y-1)^2 \Big(B^2 r_+^2
   \sin^{2}{\theta} +1\Big){}^2 \Big(B^2 r_+^3 \sin^{2}{\theta}  \Big(B^2 r_+ \Big(2 r_+-3 M\Big) \sin^{2}{\theta} +2\Big)+M\Big)\Big)\nonumber\\&\times&\Big(\pi ^3
   r_+^4\Big)^{-1}\Big)\Big)+M\Big)\Big)\Big(\pi ^2 r_+^4\Big)^{-1}.
\end{eqnarray}
\begin{center}
\includegraphics[width=8cm]{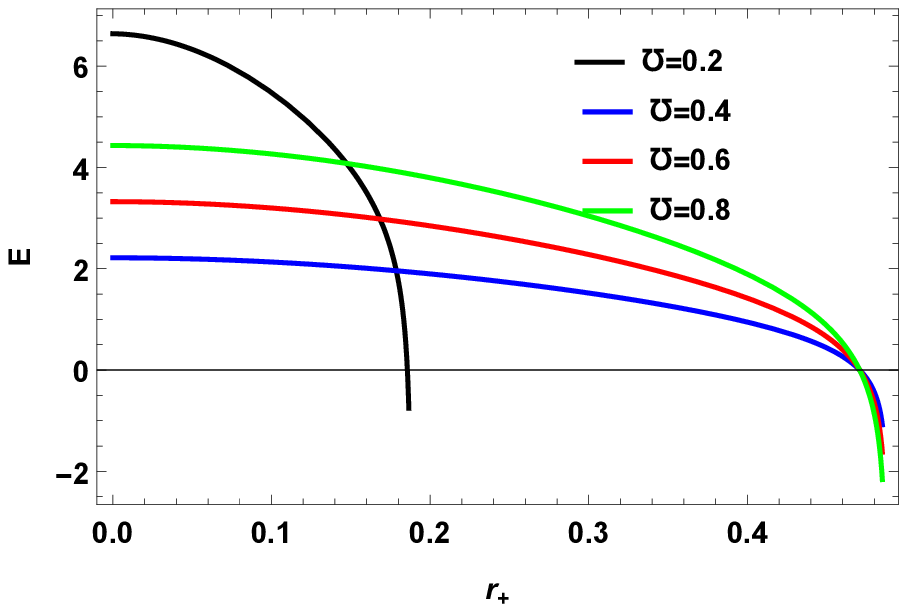}\\
{Figure 4: $E$ versus $r_{+}$ for  $M=10,~B=0.5, \alpha=0.2$  and $Y=0.5$}.
\\
\end{center}
The graph behaviour of internal energy for the different choices of horizon radius is shows in \textbf{Fig. 4}. It is noted that for the small values of radii, the graph is gradually decreases even shifts towards negative side, While the corrected internal energy depicts positive behaviour. This mean that the considered BH absorbing more and more heat from the surrounding to maintain its state. Since, BHs considered as a thermodynamic system, so there is another important thermodynamic quantity that is pressure. In this regard, there is deep connection between thermodynamic voulme  and pressure. The Expression of BH pressure ($P=-\frac{dF}{dV}$) under the effect of thermal fluctuations takes the form \textit{Appendix~B} in Eq. (\ref{RA1}).

 \begin{center}
\includegraphics[width=8cm]{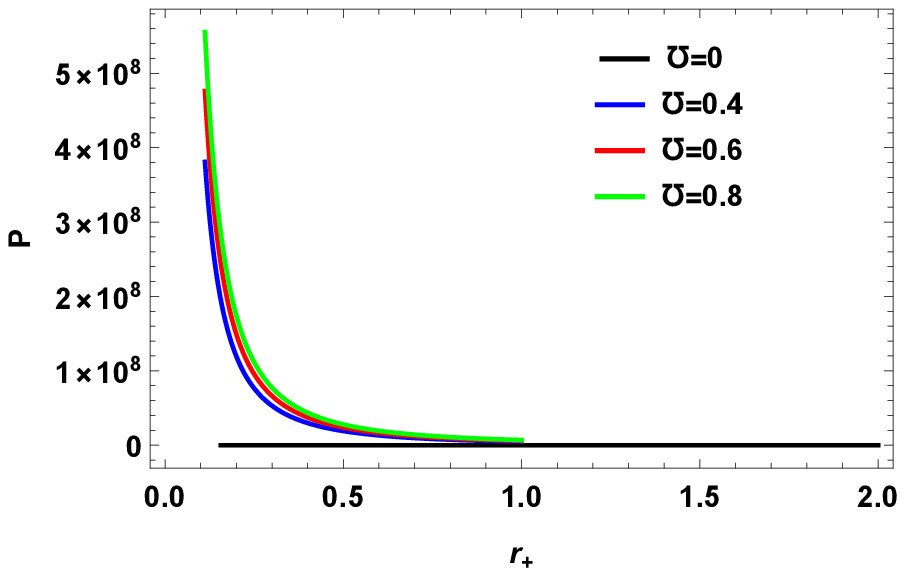}\\
{Figure 5: $P$ versus $r_{+}$ for  $M=10,~B=0.5, \alpha=0.2$  and $Y=0.5$}.
\\
\end{center}
In the \textbf{Fig. 5}, the graph of pressure is just coincides the state of equilibrium. For the different values of correction parameter, the pressure is significantly increases for the considered system. Further, there is another important thermodynamic quantity enthalpy ($H=E+PV$) is given as
\begin{eqnarray}
H&=&\Big(2 B^2 M \sin^{2}{\theta}  (\alpha  Y-1) \Big(r_+ \Big(B^2 r_+^2 \sin^{2}{\theta} -1\Big) \Big(2 B^2 r_+ \sin^{2}{\theta}  \Big(B^2 r_+ \Big(6 M-5
   r_+\Big) \sin^{2}{\theta} -3\Big) \Big(8 \pi  B^4 M^2 r_+^2 \mho  \sin^{2}{\theta} ^2\nonumber\\&\times& (\alpha  Y-1)^2 \Big(B^2 r_+^2 \sin^{2}{\theta} +1\Big){}^2-1\Big)-4
   \pi  \mho  \log \Big(\Big(8 B^4 M^2 \sin^{2}{\theta} ^2 (\alpha  Y-1)^2 \Big(B^2 r_+^2 \sin^{2}{\theta} +1\Big){}^2 \Big(B^2 r_+^3 \sin^{2}{\theta}  \Big(B^2
   r_+ \nonumber\\&\times& \Big(2 r_+-3 M\Big) \sin^{2}{\theta} +2\Big)+M\Big)\Big)\Big(\pi ^3 r_+^4\Big)^{-1}\Big)\Big)-\Big(r_+ \Big(B^2 r_+ \Big(r_++1\Big) \ \sin^{2}{\theta}+1\Big)-1\Big) \Big(r_+^2 \Big(B^2 r_+ \sin^{2}{\theta}  \Big(B^2 r_+ \nonumber\\&\times& \Big(2 r_+-3 M\Big) \sin^{2}{\theta} +2\Big)-2 \pi  \mho  \log \Big(\Big(8 B^4 M^2
   \sin^{2}{\theta} ^2 (\alpha  Y-1)^2 \Big(B^2 r_+^2 \sin^{2}{\theta} +1\Big){}^2 \Big(B^2 r_+^3 \sin^{2}{\theta}  \Big(B^2 r_+ \Big(2 r_+-3 M\Big)\nonumber\\&\times& 
   \sin^{2}{\theta} +2\Big)+M\Big)\Big)\Big(\pi ^3 r_+^4\Big)^{-1}\Big)\Big)+M\Big)\Big)\Big)\Big(\pi ^2 r_+^4\Big)^{-1}.
\end{eqnarray}
   \begin{center}
\includegraphics[width=8cm]{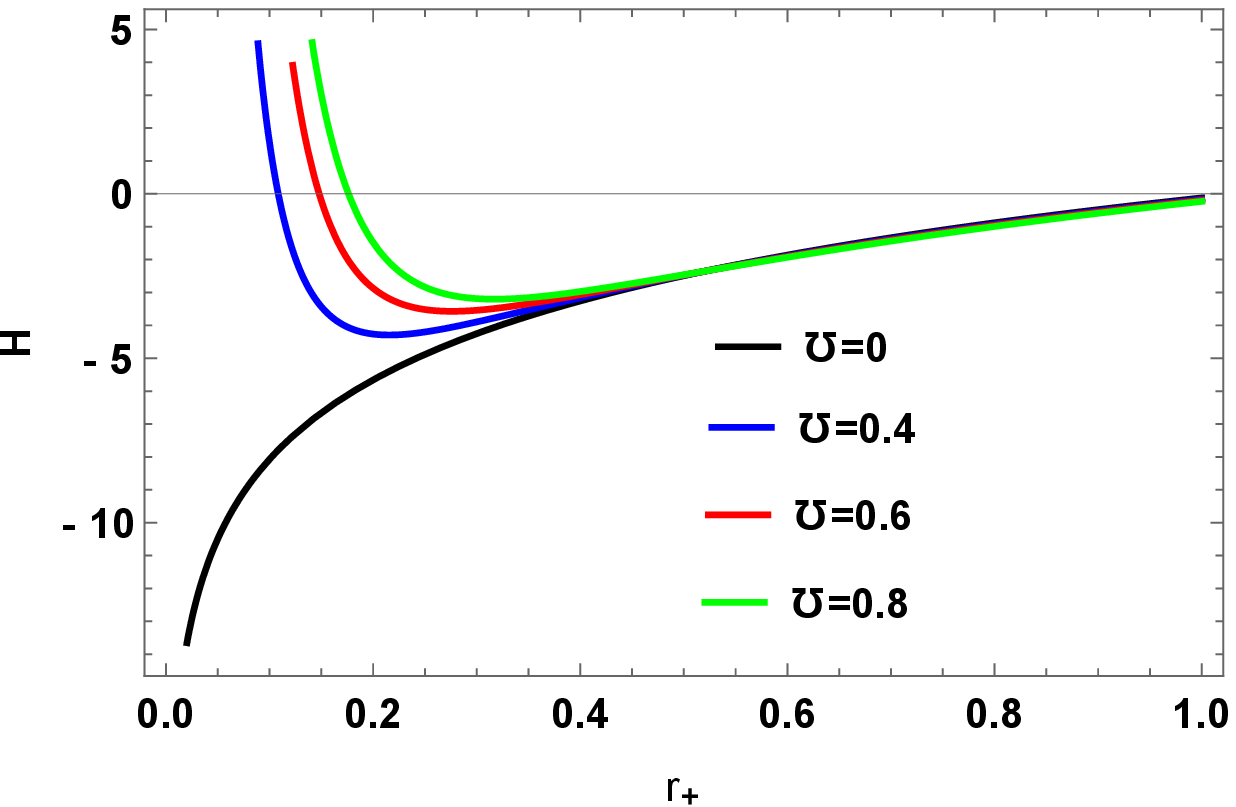}\\
{Figure 6: $H$ versus $r_{+}$ for  $M=10,~B=0.5, \alpha=0.2$  and $Y=0.5$}.
\\
\end{center}
From \textbf{Fig. 6}, it can observed that the graph of usual enthalpy is coincide with the plots of corrected one and abruptly decreases even shifts towards negative side. This means that there exists a exothermic reactions means there will be huge amount of energy release into its surroundings. In the presence of thermal fluctuations, the Gibbs free energy ($G=H-T\tilde{S}$) is expressed  \textit{Appendix~B} in Eq. (\ref{RA}).
  \begin{center}
\includegraphics[width=8cm]{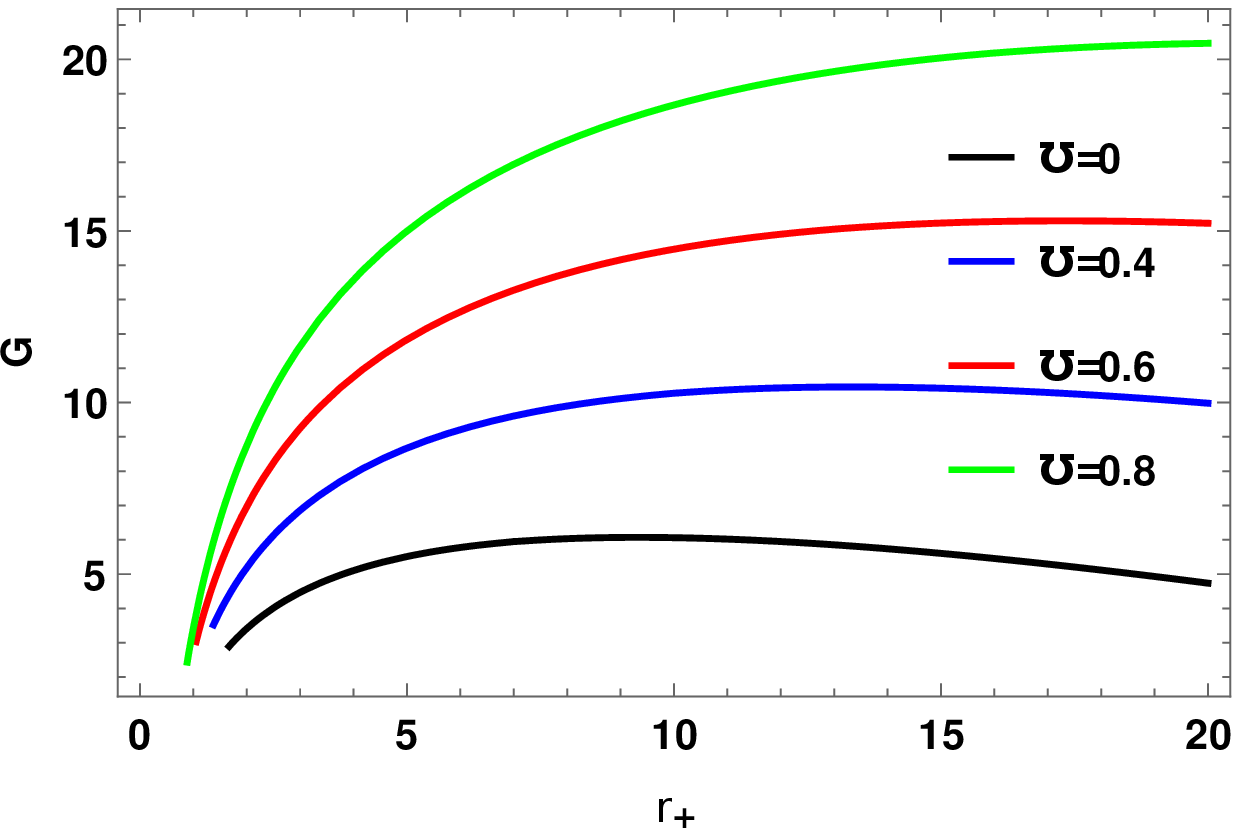}\\
{Figure 7: $G$ versus $r_{+}$ for  $M=10,~B=0.5, \alpha=0.2$  and $Y=0.5$}.
\\
\end{center}
The graph analysis of Gibbs free energy with respect to horizon radius is shows in \textbf{Fig. 7}. The positivity of this energy is sign of occurrence of non-spontaneous reactions means this system requires more energy to gain equilibrium state. After the detail discussion of thermodynamics quantities, there is another important concept is the stability of the system that is checked by specific heat. The specific heat  ($C_{\tilde{S}}=\frac{dE}{dT}$) is given as
\begin{eqnarray}
C_{\tilde{S}}&=&\Big(-r_+ \Big(r_+ \Big(B^2 r_+ \Big(r_++1\Big) \sin^{2}{\theta} +1\Big)-1\Big)+4 \Big(r_+ \Big(B^2 r_+ \Big(r_++1\Big) \sin^{2}{\theta} +1\Big)-1\Big)
   \Big(r_+^2 \Big(B^2 r_+ \sin^{2}{\theta} \nonumber\\&\times&\ \Big(B^2 r_+ \Big(2 r_+-3 M\Big) \sin^{2}{\theta} +2\Big)-2 \pi  \mho  \log \Big(\Big(8 B^4 M^2  \sin^{2}{\theta} (\alpha  Y-1)^2 \Big(B^2 r_+^2 \sin^{2}{\theta} +1\Big){}^2 \Big(B^2 r_+^3 \sin^{2}{\theta}  \Big(B^2 r_+ \nonumber\\&\times&\Big(2 r_+-3 M\Big)  \sin^{2}{\theta}+2\Big)+M\Big)\Big)\Big(\pi ^3 r_+^4\Big)^{-1}\Big)\Big)+M\Big)-r_+ \Big(B^2 r_+ \Big(3 r_++2\Big) \sin^{2}{\theta} +1\Big) \Big(r_+^2 \Big(B^2 r_+ \sin^{2}{\theta} 
   \nonumber\\&\times&\Big(B^2 r_+ \Big(2 r_+-3 M\Big) \sin^{2}{\theta} +2\Big)-2 \pi  \mho  \log \Big(\Big(8 B^4 M^2 \sin^{2}{\theta} ^2 (\alpha  Y-1)^2 \Big(B^2 r_+^2
   \sin^{2}{\theta} +1\Big){}^2 \Big(B^2 r_+^3 \sin^{2}{\theta}  \Big(B^2 r_+ \nonumber\\&\times&\Big(2 r_+-3 M\Big) \sin^{2}{\theta} +2\Big)+M\Big)\Big)\Big(\pi ^3
   r_+^4\Big)^{-1}\Big)\Big)+M\Big)\Big)\Big(2 \pi  r_+^3 \Big(B^2 r_+^2 \sin^{2}{\theta} -1\Big)\Big)^{-1}.
\end{eqnarray}
\begin{center}
\includegraphics[width=8cm]{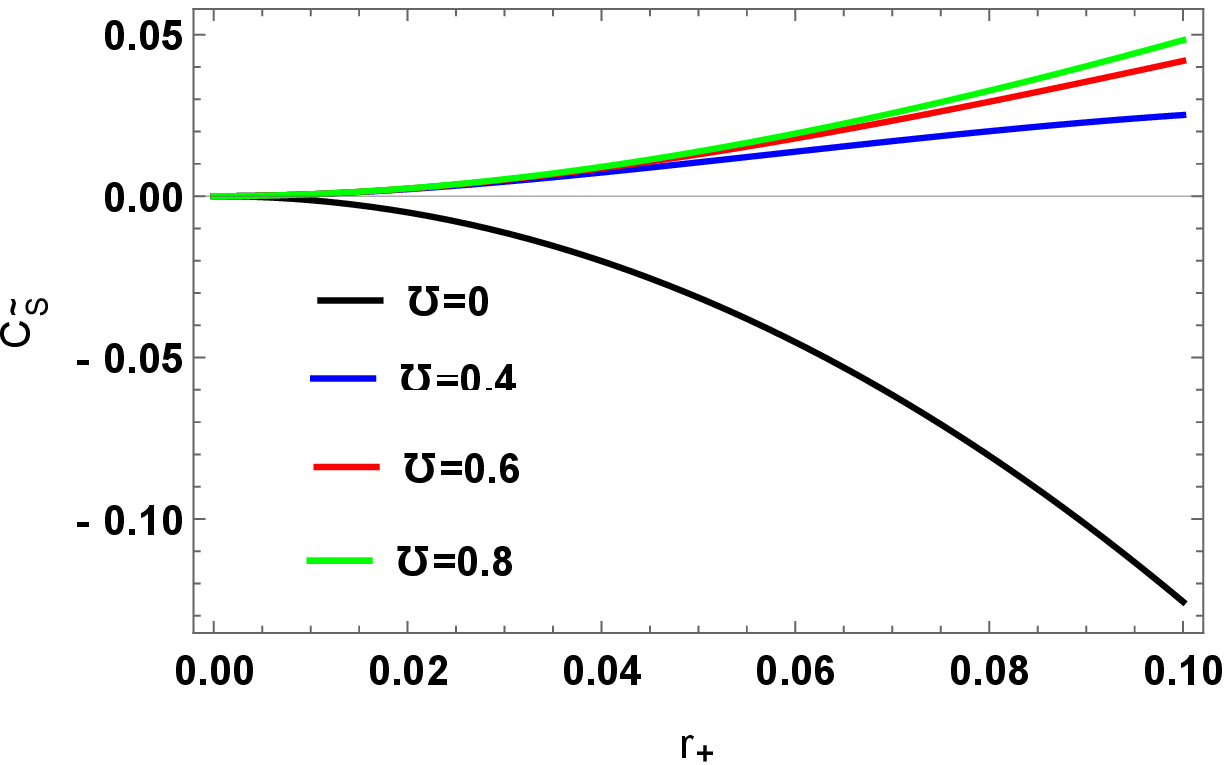}\\
{Figure 8: $C_{\tilde{S}}$ versus $r_{+}$ for $M=10,~B=0.5, \alpha=0.2$ and $Y=0.5$}.
\\
\end{center}
From {Fig. 8}, the behaviour of specific heat is observed with respect to horizon radius and different choices of correction parameter $\mho$. It can be observed that the uncorrected quantity (black) depicts negative behaviour means the system is unstable while the corrected specific heat shows positive behaviour throughout the considered domain. The positivity of this plot is indication of stable region.
\section{Discussion}

In this article, we examined the quantum gravity impacts on the particle tunneling from a magnetic~Ernst-like BH.
In order to do this, we developed a modified Lagrangian equation that included quantum effects to explain 
the motion of spin-1 particles while taking into consideration the GUP effects.
We determined the tunneling rates of vector bosons that use the Hamilton-Jacobi approach.
We have also evaluated this BH modified Hawking temperature. We have computed the modified tunneling 
probabilities depending on the properties of the emitted bosons, such as their energy, 
potential, surface gravity, magnetic field, total angular momentum, as well as quantum gravity.
It's also important to note that the modified tunneling probabilities and Hawking temperature are dependent on the quantum particles that tunnel as particles (BH's energy carriers) to generate 
gravity radiation. According to our analysis, the quantum correction parameter increased 
and then the temperature decreased during the radiation process.

Moreover, we have studied the effects of thermal fluctuations and logarithmic corrections via graphical analysis on the thermodynamics of given BH e.g., Hawking temperature, entropy, internal energy, heat capacity, specific heat, Helmholtz free energy, enthalpy, Gibbs free energy and pressure. It is noted that the corrected entropy is monotonically increasing throughout the considered domain and the graph of usual entropy is also increasing just for small value of horizon radius. Thus, these corrections terms are more effective for small BHs. While the behaviour of Helmholtz free energy means that the considered system shifts its state towards equilibrium and no more work can be extract from it. Moreover, the internal energy internal energy depicts positive behaviour for the choices of correction parameter, so the considered BH absorbing more and more heat from the surrounding to maintain its state. The behaviour of pressure and enthalpy depicts almost same behaviour along horizon radius, like it can observed that the graph of usual enthalpy is coincide with the plots of corrected one and abruptly decreases even shifts towards negative side. This means that there exists a exothermic reactions means there will be huge amount of energy release into its surroundings. Finally, to check the stability of the system, the behaviour of specific heat is observed with respect to horizon radius and different choices of correction parameter $\mho$. It can be concluded that these correction terms makes the system stable under thermal fluctuations.\\

\section*{Acknowledgments}
The work of KB was supported in part by the JSPS KAKENHI Grant Number JP21K03547.

\section*{Appendix A}
The set of field equations is obtained by applying the WKB approximation to the Lagrangian equation.
\begin{eqnarray}
&&\frac{1}{G}[c_{1}(\partial_{0}K_{0})
(\partial_{1}K_{0})+c_{1}\alpha(\partial_{0}K_{0})^{3}(\partial_{1}K_{0})
-c_{0}(\partial_{1}K_{0})^{2}-c_{0}(\partial_{1}K_{0})^{4}\alpha+e A_{0}c_{1}(\partial_{1}K_{0})
+e A_{0}c_{1}\alpha(\partial_{1}K_{0})(\partial_{0}K_{0})^{2}]\nonumber\\
&&+\frac{1}{H}
[c_{2}(\partial_{0}K_{0})(\partial_{2}K_{0})
+\alpha c_{2}(\partial_{0}K_{0})^{3}(\partial_{2}K_{0})
-c_{0}(\partial_{2}K_{0})^{2}-\alpha c_{0}(\partial_{2}K_{0})^{4}
+eA_{0}c_{2}(\partial_{2}K_{0})+
\alpha eA_{0}c_{2}(\partial_{0}K_{0})^{2}
(\partial_{2}K_{0})]\nonumber\\
&&+\frac{1}{I}
[c_{3}(\partial_{0}K_{0})(\partial_{3}K_{0})
+\alpha c_{3}(\partial_{0}K_{0})^{3}(\partial_{3}K_{0})+
c_{0}(\partial_{3}K_{0})^{2}
+\alpha c_{0}(\partial_{3}K_{0})^{4}
+eA_{0}c_{3}(\partial_{3}K_{0})+\alpha c_{3}eA_{0}
(\partial_{0}K_{0})^{2}(\partial_{3}K_{0})]\nonumber\\
&&-m^{2}c_{0}=0 \label{31}
\end{eqnarray}
\begin{eqnarray}
&&\frac{-1}{F}[c_{0}(\partial_{0}K_{0})(\partial_{1}K_{0})
+c_{0}\alpha (\partial_{0}K_{0})(\partial_{1}K_{0})^{3}
-c_{1}(\partial_{0}K_{0})^{2}-
c_{1}\alpha (\partial_{0}K_{0})^{4}-eA_{0}c_{1}(\partial_{0}K_{0})-\alpha
eA_{0}c_{1}(\partial_{1}K_{0})^{2}(\partial_{0}K_{0})]\nonumber\\
&&
+\frac{1}{H}[c_{2}(\partial_{1}K_{0})(\partial_{2}K_{0})+\alpha c_{2}(\partial_{1}K_{0})^{3}(\partial_{2}K_{0})
-c_{1}(\partial_{2}K_{0})^{2}-\alpha c_{1}(\partial_{2}K_{0})^{4}]
+\frac{1}{I}[c_{3}(\partial_{1}K_{0})
(\partial_{3}K_{0})+c_{3}\alpha
(\partial_{1}K_{0})^{3}
(\partial_{3}K_{0})\nonumber\\
&&-c_{1}(\partial_{3}S_{0 })^{2}-c_{1}\alpha
(\partial_{3}K_{0})^{4}]
-m^{2}c_{1}-\frac{1}{F}eA_{0}[c_{0}
(\partial_{1}K_{0})+\alpha c_{0}(\partial_{1}K_{0})^{3}-c_{1}(\partial_{0}K_{0})
-\alpha c_{1}(\partial_{0}K_{0})^{3}-c_{1}eA_{0}
\nonumber\\
&&-eA_{0}\alpha c_{1}
(\partial_{1}K_{0})^{2}]
=0\label{41}\\
&&\frac{1}{F}[c_{0}(\partial_{0}K_{0})(\partial_{2}K_{0})
+\alpha c_{0}(\partial_{0}K_{0})(\partial_{2}K_{0})^{3}
-c_{2}(\partial_{0}K_{0})^{2}-\alpha c_{2}(\partial_{0}K_{0})^{4}
-eA_{0}(\partial_{0}K_{0})c_{2}
-eA_{0}(\partial_{0}K_{0})^{3}c_{2}\alpha
]-\frac{1}{G}\nonumber\\
&&[c_{2}
(\partial_{1}K_{0})^{2}+\alpha c_{2}(\partial_{1}K_{0})^{4}
-c_{1}(\partial_{1}K_{0})(\partial_{2}K_{0})-\alpha c_{1}
(\partial_{1}K_{0})(\partial_{2}K_{0})^{3}]+\frac{1}{I}
[c_{3}(\partial_{2}K_{0})(\partial_{3}K_{0})
+\alpha c_{3}(\partial_{2}K_{0})^{3}(\partial_{3}K_{0})
\nonumber\\
&&-c_{2}(\partial_{3}K_{0})^{2}-
\alpha (\partial_{3}K_{0})^{4}c_{2}]
-\frac{A_{0}e}{F}[(\partial_{2}K_{0})c_{0}
+\alpha (\partial_{2}K_{0})^{3}c_{0}-
(\partial_{0}K_{0})c_{2}-\alpha (\partial_{0}K_{0})^{3}c_{2}
+eA_{0}c_{2}+\alpha c_{2}eA_{0}(\partial_{0}K_{0})^{2}]\nonumber\\
&&-c_{2}m^{2}=0 \label{51}\\
&&\frac{1}{F}[(\partial_{0}K_{0})(\partial_{3}K_{0})c_{0}
+\alpha (\partial_{0}K_{0})(\partial_{3}K_{0})^{3}c_{0}
-(\partial_{0}K_{0})^{2}c_{3}-\alpha (\partial_{0}K_{0})^{4}c_{3}
-eA_{0}(\partial_{0}K_{0})c_{3}
-eA_{0}(\partial_{3}K_{0})^{2}(\partial_{0}
K_{0})c_{3}\alpha]\nonumber\\
&&+\frac{1}{G}[c_{3}
(\partial_{1}K_{0})^{2}+\alpha c_{3}(\partial_{1}K_{0})^{4}
-c_{1}(\partial_{3}K_{0})(\partial_{1}K_{0})-\alpha c_{1}
(\partial_{1}K_{0})(\partial_{3}K_{0})^{3}]+\frac{1}{H}
[c_{3}(\partial_{2}K_{0})^{2}+\alpha c_{3}(\partial_{2}K_{0})^{4}
-c_{2}(\partial_{2}K_{0})\nonumber\\
&&(\partial_{3}K_{0})-
\alpha c_{2}(\partial_{3}K_{0})^{3}(\partial_{2}K_{0})]
+\frac{eA_{0}}{F}[c_{0}(\partial_{3}K_{0})
+\alpha c_{0}(\partial_{3}K_{0})^{3}-c_{3}
(\partial_{0}K_{0})-\alpha c_{3}(\partial_{0}K_{0})^{3}
-c_{3}eA_{0}-\alpha c_{3}eA_{0}(\partial_{3}K_{0})^{2}]\nonumber\\
&&-c_{3}m^{2}=0.\label{61}
\end{eqnarray}\\
\textbf{Appendix B}
The expression of BH pressure under the effect of thermal fluctuations takes the form,
\begin{eqnarray}
P&=&\Big(2 B^2 M \sin^{2}{\theta}  (\alpha  Y-1) \Big(B^2 r_+^2 \sin^{2}{\theta} -1\Big) \Big(2 r_+ \Big(B^2 r_+ \sin^{2}{\theta}  \Big(B^2 r_+ \Big(2 r_+-3
   M\Big) \sin^{2}{\theta} +2\Big)-2 \pi  \mho  \log \nonumber\\&\times& \Big(\Big(8 B^4 M^2 \sin^{2}{\theta} ^2 (\alpha  Y-1)^2 \Big(B^2 r_+^2 \sin^{2}{\theta} +1\Big){}^2
   \Big(B^2 r_+^3 \sin^{2}{\theta}  \Big(B^2 r_+ \Big(2 r_+-3 M\Big) \sin^{2}{\theta} +2\Big)+M\Big)\Big)\nonumber\\&\times&\Big(\pi ^3 r_+^4\Big)^{-1}\Big)\Big)+r_+^2 \Big(B^2 r_+
   \sin^{2}{\theta}  \Big(B^2 \Big(2 r_+-3 M\Big) \sin^{2}{\theta} +2 B^2 r_+ \sin^{2}{\theta} \Big)+B^2 \sin^{2}{\theta}  \Big(B^2 r_+ \Big(2 r_+-3
   M\Big) \sin^{2}{\theta}\Big)\nonumber\\&-&\pi ^4 r_+^4 \mho  \Big(\Big(32 B^6 M^2 \sin^{2}{\theta} ^3 (\alpha  Y-1)^2 \Big(B^2 r_+^2 \sin^{2}{\theta} +1\Big)
   \Big(B^2 r_+^3 \sin^{2}{\theta}  \Big(B^2 r_+ \Big(2 r_+-3 M\Big) \sin^{2}{\theta} +2\Big)+M\Big)\Big)\nonumber\\&\times&\Big(\pi ^3 r_+^3\Big)^{-1}+\Big(8 B^4 M^2 \sin^{2}{\theta} ^2
   (\alpha  Y-1)^2 \Big(B^2 r_+^2 \sin^{2}{\theta} +1\Big){}^2 \Big(B^2 r_+^3 \sin^{2}{\theta}  \Big(B^2 \Big(2 r_+-3 M\Big) \sin^{2}{\theta} +2 B^2 r_+
   \sin^{2}{\theta} \Big)\nonumber\\&+&3 B^2 r_+^2 \sin^{2}{\theta}  \Big(B^2 r_+ \Big(2 r_+-3 M\Big) \sin^{2}{\theta} +2\Big)\Big)\Big)\Big(\pi ^3 r_+^4\Big)^{-1}.\label{RA1}
\end{eqnarray}
In the presence of thermal fluctuations, the Gibbs free energy is expressed as
\begin{eqnarray}
G&=&\Big(2 B^2 M \sin^{2}{\theta}  \Big(r_+ (\alpha  Y-1) \Big(B^2 r_+^2 \sin^{2}{\theta} +1\Big) \Big(B^2 r_+^3 \sin^{2}{\theta}  \Big(B^2 r_+ \Big(2 r_+-3
M\Big) \sin^{2}{\theta} +2\Big)-2 \pi  r_+^2 \mho  \log \nonumber\\&\times&\Big(\Big(8 B^4 M^2 \sin^{2}{\theta} ^2 (\alpha  Y-1)^2 \Big(B^2 r_+^2 \sin^{2}{\theta} +1\Big){}^2
 \Big(B^2 r_+^3 \sin^{2}{\theta}  \Big(B^2 r_+ \Big(2 r_+-3 M\Big) \sin^{2}{\theta} +2\Big)+M\Big)\Big)\Big(\pi ^3 r_+^4\Big)^{-1}\Big)\Big)\nonumber\\&+&(\alpha  Y-1) \Big(r_+
   \Big(B^2 r_+^2 \sin^{2}{\theta} -1\Big) \Big(2 B^2 r_+ \sin^{2}{\theta}  \Big(B^2 r_+ \Big(6 M-5 r_+\Big) \sin^{2}{\theta} -3\Big) \Big(8 \pi  B^4 M^2
   r_+^2 \mho  \sin^{2}{\theta} ^2 (\alpha  Y-1)^2\nonumber\\&\times& \Big(B^2 r_+^2 \sin^{2}{\theta} +1\Big){}^2-1\Big)-4 \pi  \mho  \log \Big(\Big(8 B^4 M^2 \sin^{2}{\theta} ^2
   (\alpha  Y-1)^2 \Big(B^2 r_+^2 \sin^{2}{\theta} +1\Big){}^2 \Big(B^2 r_+^3 \sin^{2}{\theta}  \Big(B^2 r_+ \Big(2 r_+-3 M\Big)\nonumber\\&\times& \sin^{2}{\theta} +2\Big)+M\Big)\Big)\Big(\pi ^3 r_+^4\Big)^{-1}\Big)\Big)-\Big(r_+ \Big(B^2 r_+ \Big(r_++1\Big) \sin^{2}{\theta} +1\Big)-1\Big) \Big(r_+^2 \Big(B^2 r_+
   \sin^{2}{\theta}  \Big(B^2 r_+ \Big(2 r_+-3 M\Big) \nonumber\\&\times&\sin^{2}{\theta} +2\Big)-2 \pi  \mho  \log \Big(\Big(8 B^4 M^2 \sin^{2}{\theta} ^2 (\alpha  Y-1)^2
   \Big(B^2 r_+^2 \sin^{2}{\theta} +1\Big){}^2 \Big(B^2 r_+^3 \sin^{2}{\theta}  \Big(B^2 r_+ \Big(2 r_+-3 M\Big) \sin^{2}{\theta}+2\Big)\Big)\Big)\nonumber\\&\times&\Big(\pi ^3
r_+^4\Big)^{-1}\Big)\Big)+M\Big)\Big)\Big)\Big)\Big(\pi ^2 r_+^4\Big)^{-1}.\label{RA}
\end{eqnarray}

\section*{Data availability}
The data for this manuscript consists of References in the literature. All references can be found in academic libraries.


\begin{thebibliography}{99}

\bibitem{1} S. W. Hawking, Nature \textbf{248}, 30(1974).

\bibitem{2} R. Kerner, R. B. Mann, Class. Quant. Grav. \textbf{25}, 095014(2008).
ibid; Phys. Rev. \textbf{D 73}, 104010(2006).

\bibitem{3} Q. Q. Jiang, S. Q. Wu, X. Cai, Phys. Rev. \textbf{D 73}, 064003(2006).

\bibitem{4} H. Erbin, V. Lahoche, Phys. Rev. \textbf{D 98}, 104001(2018).

\bibitem{5} M. K. Parikh, F. Wilczek, Phys. Rev. Lett. \textbf{85}, 5042(2000).

\bibitem{6} K. Srinivasan, T. Padmanabhan, Phys. Rev. \textbf{D 60}, 024007(1999).

\bibitem{7} R. Banerjee, B. R. Majhi, J. High Energy Phys. \textbf{06}, 095(2008).

\bibitem{8} M. K. Parikh, F. Wilczek, Phys. Rev. Lett. \textbf{85}, 5042(2000).

\bibitem{9} K. Srinivasan, T. Padmanabhan, Phys. Rev. \textbf{D 60}, 024007(1999).

\bibitem{10} L. Brillouin, Comptes Rendus de l?Academie des Sciences \textbf{183}, 24(1926).

\bibitem{11}  I. Sakalli and A. \"{O}vg\"{u}n, Gen. Rel. Grav.  {\bf 48}, No. 1, 1 (2016).

\bibitem{12} I. Sakalli and A. \"{O}vg\"{u}n, Eur. Phys. J. Plus {\bf 130}, No. 6, 110 (2015).

\bibitem{15} I. Sakalli and A. \"{O}vg\"{u}n, Astrophys. Space Sci.  {\bf 359}, No. 1, 32 (2015).

\bibitem{21} X. M. Kuang, J. Saavedra and A. \"{O}vg\"{u}n, Eur. Phys. J. C {\bf 77}, No. 9, 613 (2017).

\bibitem{22} X. M. Kuang, B. Liu and A. \"{O}vg\"{u}n, Eur. Phys. J. C {\bf 78}, No. 10, 840 (2018).

\bibitem{23} W. Javed, R. Babar and A. \"{O}vg\"{u}n, Mod. Phys. Lett. A {\bf 34}, No. 09, 1950057 (2019).

\bibitem{24} P. A. Gonzalez, A. \"{O}vg\"{u}n, J. Saavedra and Y. Vasquez, Gen. Rel. Grav.  {\bf 50}, No. 6, 62 (2018).

\bibitem{27} Deyou Chen, Houwen Wu, Haitang Yang, Shuzheng Yang, Int. J. Mod. Phys. \textbf{A 29} (2014) 1430054.

\bibitem{28} Deyou Chen, Eur. Phys. J. \textbf{C 74} (2014) 2687.

\bibitem{29} G. Gecim and Y. Sucu, Mod. Phys. Lett. A {\bf 33}, No. 28, 1850164 (2018).

\bibitem{30} G. Gecim and Y. Sucu, Phys. Lett. B {\bf 773}, 391 (2017).

\bibitem{32} W. Javed, R. Ali, R. Babar, A. \"{O}vg\"{u}n, Eur. Phys. J. Plus {\bf 134}, 511 (2019).

\bibitem{33} W. Javed, R. Ali, R. Babar, A. \"{O}vg\"{u}n, Chinese Physics {\bf C 144}, 015104(2020).

\bibitem{35} M. Hossain Ali, Class. Quant. Grav. \textbf{24}, 5849 (2007).

\bibitem{37} M. Hossain Ali, Gen. Rel. Grav. \textbf{36}, 1171 (2004).

\bibitem{39}  E. T. Akhmedov, V. Akhmedova and D. Singleton, Phys. Lett. B \textbf{642}, 124 (2006).

\bibitem{42} T. Zhu, J. R. Ren and D. Singleton, Int. J. Mod. Phys. D {\bf 19}, 159 (2010).

\bibitem{43} V. Akhmedova, T. Pilling, A. de Gill and D. Singleton, Phys. Lett. B {\bf 666}, 269 (2008).

\bibitem{44} V. Akhmedova, T. Pilling, A. de Gill and D. Singleton, Phys. Lett. B {\bf 673}, 227 (2009).

\bibitem{45} E. T. Akhmedov, V. Akhmedova, T. Pilling and D. Singleton, Int. J. Mod. Phys.\ A {\bf 22}, 1705 (2007).

\bibitem{46} G. Gecim and Y. Sucu, Gen. Rel. Grav. {\bf 50}, No. 12, 152 (2018).

\bibitem{47} I. A. Meitei, T. I. Singh, S. G. Devi, N. P. Devi and K. Y. Singh, Int. J. Mod. Phys. A {\bf 33}, No. 12, 1850070 (2018).

\bibitem{48} G. R. Chen, S. Zhou and Y. C. Huang, Int. J. Mod. Phys. D {\bf 24}, No. 01, 1550005 (2014).

\bibitem{49} X. X. Zeng and S. Z. Yang, Chin. Phys. B {\bf 18}, 462 (2009).

\bibitem{50} S. K. Modak, Phys. Lett. B {\bf 671}, 167 (2009).

\bibitem{51} R. Li and J. R. Ren, Phys. Lett. B {\bf 661}, 370 (2008).

\bibitem{53} Q. Q. Jiang, S. Q. Wu and X. Cai, Phys. Lett. B {\bf 651}, 58 (2007).

\bibitem{54} X. He and W. Liu, Phys. Lett. B {\bf 653}, 330 (2007).

\bibitem{56} H. L. Li, S. Z. Yang, Q. Q. Jiang and D. J. Qi, Phys. Lett. B {\bf 641}, 139 (2006).

\bibitem{57} X. Q. Li, G.R. Chen, Phys. Lett. B {\bf 751}, 34(2015).

\bibitem{58} G. R. Chen, Y.C. Huang, Int. J. Mod. Phys. Rev. A {\bf 30}, 1550083(2015).

\bibitem{59} Z. Feng, Y. Chen, X. Zu, Astrophys. space Sci. {\bf 48}, 359(2015).

\bibitem{60} K. Jusufi, Ali \"{O}vg\"{u}n, Astrophys. Space Sci. {\bf 361}, 207(2016).

\bibitem{61} W. Javed, G. Abbas, R. Ali, Eur. Phys. J. C {\bf 77}(2017)296.

\bibitem{63} W. Javed, R. Ali, G. Abbas, Can. J. Phys. {\bf 97}, 176(2018).

\bibitem{64} T. Jian, C. Bing-Bing, Acta Physica Polonica, {\bf 40}, 241(2009).

\bibitem{65} A. Yale, Phys. Lett. B {\bf 697}, 398(2011).

\bibitem{66} M. Sharif, W. Javed, Can. J. Phys. \textbf{90}, 903(2012); \textit{ibid}.
Gen. Relativ. Gravit. \textbf{45}, 1051(2013).

\bibitem{67} M. Sharif, W. Javed, Can. J. Phys. {\bf 91}, 43(2013).

\bibitem{68} M. Sharif, W. Javed, Eur. Phys. J. C {\bf 72}, 1997(2012).

\bibitem{68a} R. Ali, K. Bamba, S. A. A. Shah and M. J. Saleem, Int. J. Mod. Phys. D {\bf 31},No. 09, 2250069(2022).

\bibitem{69} J. M. Bardeen, in Conference Proceedings of GR5 (Tbilisi, URSS, 1968), p. 174.

\bibitem{70} D. Chen, H. Wu and H. Yang, Adv. High Energy Phys.  {\bf 2013}, 432412 (2013).

\bibitem{71} A. D. Vries, T. S. Kaler, Phys. Rev. D {\bf 65}, 104022(2002).

\bibitem{72} Q. Q. Jiang, Phys. Rev. D {\bf 78}, 044009(2008).

\bibitem{78} R. Ali, R. Babar and M. Asgher, Annalen der Physik, {\bf 2200074}, 12(2022).

\bibitem{82} K. Ghaderi, B. Malakolkalami, Nucl. Phys. \textbf{B 903}, 10(2016).

\bibitem{T4} R. Ali, M. Asgher, New Astronomy {\bf 93}, 101759(2022).

\bibitem{T5} R. Ali, R. Babar, P. K. Sahoo, Physics of the Dark Universe {\bf 35}, 100948(2022).

\bibitem{T7}  M. Sharif, and  Z. Akhtar, Phys. Dark Universe {\bf 29}, 100589(2020).

\bibitem{T8}  M. Sharif, and  Z. Akhtar, Chin. J. Phys {\bf 71}, 669(2021).

\bibitem{T9} W, Javed, Z. Yousaf, Z. Akhtar, Mod. Phys. Lett. {\bf A 33}, 1850089(2018).

\bibitem{T10} Z, Yousaf, K. Bamba, Z. Akhtar and W. Javed, Int. J. Geom. Methods Mod. {\bf 19}, 2250102(2022).

\bibitem{c1} R. A. Konoplya, R.D.B. Fontana, Phys. Lett. {\bf B 659}, 375(2008).

\bibitem{c2} S. Shaymatov, M. Jamil, K. Jusufi and K. Bamba,  Eur. Phys. J. {\bf C 82}, 636(2022).
\end{thebibliography}
\end{document}